\documentclass[a4paper,11pt]{emulateapj}
\usepackage{amssymb, latexsym}
\usepackage{graphicx}
\usepackage{longtable}
\usepackage{amsmath} 

\renewcommand\ion[2]{#1$\;${\scshape{#2}}}

\def\oii{\ion{O}{ii}}
\def\nii{\ion{N}{ii}}
\def\sii{\ion{S}{ii}}

\def\hii{\ion{H}{ii}}

\def\oiii{\ion{O}{iii}}

\def\neiii{\ion{Ne}{iii}}

\def\sigsfr{$\Sigma_{\mathrm{SFR}}$}
\def\reff{$R_{\mathrm{eff}}^2$}

\def\ha {\mbox{{\rm H}$\alpha$}}

\def\hb {\mbox{{\rm H}$\beta$}}
\usepackage{footmisc}
\usepackage{hyperref}
\usepackage{perpage} 
\MakePerPage{footnote} 

\begin{document}
\title{A tight relation between N/O ratio and galaxy stellar mass can explain the evolution of strong emission line ratios with redshift}

\author{Daniel Masters\altaffilmark{1}, Andreas Faisst\altaffilmark{1}, Peter Capak\altaffilmark{2}}

\altaffiltext{1}{Infrared Processing and Analysis Center, Caltech,
  Pasadena, CA 91125}
\altaffiltext{2}{Spitzer Science Center, California Institute of Technology, Pasadena, CA 91125, USA}




\begin{abstract}

The offset of high redshift star-forming galaxies in the [\oiii]/\hb\ versus [\nii]/\ha\ (O3N2) diagram in comparison with  the local star-forming galaxy sequence is now well established. The physical origin of the shift is the subject of some debate, and has important implications for metallicity measurements based on strong lines at all redshifts. To investigate the origin of the O3N2 offset, we use a sample of $\sim$100,000 star-forming galaxies from SDSS DR12 to probe the empirical correlations between emission line diagnostics and measurable galaxy physical properties. In particular, we examine how surface density of star formation (\sigsfr), ionization parameter, nitrogen-to-oxygen (N/O) ratio, and stellar mass drive position in two key diagnostic diagrams: O3N2 and [\oiii]/\hb\ versus [\sii]/\ha\ (O3S2). We show that, at fixed [\oiii]/\hb, galaxies falling closer to the high-redshift locus in O3N2 have higher \sigsfr, stellar mass and N/O ratios. We also find a tight correspondence in the distributions of stellar mass and N/O in the diagnostic diagrams. This relation, spanning a range of galaxy evolutionary states, suggests that the N/O-$M_{*}$ relation is more fundamental than the N/O-metallicity relation. We argue that a tight N/O-$M_{*}$ relation is well-motivated physically, and that the observed correlation of N/O with O/H in the local universe is primarily a reflection of the existence of the mass-metallicity relation. Because the mass-metallicity relation evolves much more rapidly with redshift than N/O-$M_{*}$, the N/O ratios of high redshift galaxies are significantly elevated in comparison with local galaxies with the same gas-phase O/H. The O3N2 shift and elevated N/O ratios observed in high redshift galaxies therefore come about as a natural consequence of the N/O-$M_{*}$ relation combined with the evolution of the mass-metallicity relation.

\end{abstract}
\maketitle
\flushbottom

\section{Introduction}

Optical emission lines contain a wealth of information about the physics of the gas and stars 
in star-forming galaxies. Collisional and recombination lines encode information about the predominant ionizing radiation source, dust content, density, temperature, metallicity, and kinematics of 
the \hii\ region gas. Classical emission line diagnostic diagrams such as the [\oiii]/\hb\ vs. [\nii]/\ha\ (O3N2) \citep{Baldwin81} and  [\oiii]/\hb\ vs. [\sii]/\ha\ (O3S2)  \citep{Veilleux87} ``BPT'' diagrams are widely used to separate star-forming galaxies and active galactic nuclei (AGN) based on their different excitation mechanisms. Moreover, star-forming galaxies in the local universe are found to follow a remarkably tight sequence in these diagrams, which is generally interpreted as resulting from relationships between ionization state and gas-phase and stellar metallicity \citep{McCall85, Dopita86, Kewley01}. 

Recent observations of the rest-frame optical emission
lines of high redshift ($z\sim2-3$) galaxies 
 have revealed a pronounced shift in the location of the
ionization sequence of high-redshift galaxies with respect to local
galaxies in the O3N2 diagram \citep{Masters14, Steidel14,
  Shapley15}, confirming earlier hints at this shift from smaller samples
\citep{Shapley05, Erb06, Liu08}. The O3N2 shift  (which is in the sense of higher [\oiii]/\hb\ at fixed [\nii]/\ha\, or, alternatively, higher [\nii]/\ha\ at fixed [\oiii]/\hb) has important implications for metallicity measurement from strong emission lines at high redshift. Several commonly-used strong line metallicity indicators are calibrated against the tight locus of local galaxies in O3N2; the observed change of the locus  with redshift calls into question the validity of metallicities derived from these lines at different epochs. The physical origin of the O3N2 shift is therefore crucial to understand. 

Proposed physical causes of the shift are generally linked to the more ``extreme" star-forming conditions expected to exist at high redshift. Some of the suggested causes include higher ionization parameters in the high-redshift galaxies \citep{Brinchmann08b, Kewley15, Kashino16}, elevated nitrogen-to-oxygen (N/O) ratios at fixed O/H \citep{Masters14, Shapley15, Jones15, Sanders16, Cowie16}, and harder radiation fields due to hotter ionizing stars \citep{Steidel14}. \cite{Steidel14} suggested that there may in fact be contributions (to varying degrees) from all of these effects.

The interpretation of the O3N2 shift as being due to elevated N/O values at a fixed O/H in high redshift galaxies has gained some traction recently, but suggestions for why this might be the case have been problematic. For example, \citet{Masters14} argued that elevated N/O ratios might come from an enhanced population of Wolf-Rayet (WR) stars in high redshift galaxies. WR stars can release large amounts of nitrogen and temporarily boost the N/O ratios of galaxies in which they are found \citep{Brinchmann08a}. However, as pointed out by \citet{Shapley15}, finding a significant WR population depends on catching galaxies at an extremely young stellar age, so this explanation is unlikely the cause of the overall effect.  Motivated by stellar population models of massive star binarity (e.g., \citealp{Eldridge09}), \citet{Steidel14} argued that a higher nitrogen yield during the main-sequence evolution of massive stars could arise from an increased fraction of massive binaries at high redshift. Massive star binarity (which can lead to rapid rotation and efficient ejection of nucleosynthetic material) might plausibly be more common in the dense star-forming environments found at high redshift, but there are clearly significant uncertainties associated with this explanation as well.
 
More detailed studies of local galaxies could help shed light on this issue. It is becoming clear that a population of galaxies in the local universe exists that resembles, in most or all measurable respects, typical galaxies found at $z\sim2-3$ \citep{Brinchmann08b, Cardamone09, Stanway14, Cowie16, Bian16, Faisst16a, Greis16}. In fact, such sources can be reliably identified through their position on the O3N2 diagram \citep{Bian16}. Like their high-redshift counterparts, local galaxies lying close to the high-redshift locus in O3N2 tend to have compact sizes, high ionization parameters, and high star formation rates relative to their mass (high specific star formation rates, or sSFRs). However, they constitute an extreme tail of the local galaxy distribution and are thus quite rare. Presuming that these galaxies accurately reflect the conditions at high redshift, insight into the cause of the high-redshift BPT shift may be gained by developing a clearer sense of why they are offset in O3N2.

Here we use 104,256 star-forming galaxies from SDSS DR12  to analyze the O3N2 and O3S2 diagnostic diagrams in detail, probing the purely \emph{empirical} correlations that exist between where a galaxy falls in diagnostic diagram space and its measurable physical properties. We focus on 
four properties in particular: surface density of star formation (\sigsfr) as encoded by $L$(\ha)/\reff, nitrogen-to-oxygen abundance ratio (N/O), ionization parameter as encoded by the ratio [\neiii]/[\oii] \citep{Levesque14}, and stellar mass. The dependence of galaxy position on the O3N2 diagram on \sigsfr\ (or \hb\ luminosity, or EW(\ha), which are related) has been noted previously by a number of authors (e.g., \citealp{Brinchmann08b, Juneau14, Salim14, Cowie16}). However, the correlation itself does not address the proximate physical cause of the variation in line ratios. 

The plan of this paper is as follows. In \S2 we use the large sample of galaxy spectra from the Sloan Digital Sky Survey (SDSS) to illustrate the empirical correlations that exist between
measurable physical properties and galaxy position on the diagnostic diagrams.  In \S3 we attempt to physically interpret the observed correlations, concluding that a tight N/O versus stellar mass (N/O-$M_{*}$) relation best explains both the low- and high-redshift observations. In \S4 we address why the N/O-$M_{*}$ relation might be expected to be more fundamental than the N/O-O/H relation. In \S5 we discuss the implications of our results for metallicity measurement at different redshifts as well as for the proposed fundamental mass-metallicity-SFR (FMR) relation. We also discuss the possible origin of an observed strong correlation of \sigsfr\ with position in O3S2. In \S6 we conclude with a summary of our main results.

\begin{figure*}[htb]
\centering
  \begin{tabular}{@{}cc@{}}
    \includegraphics[width=.45\textwidth]{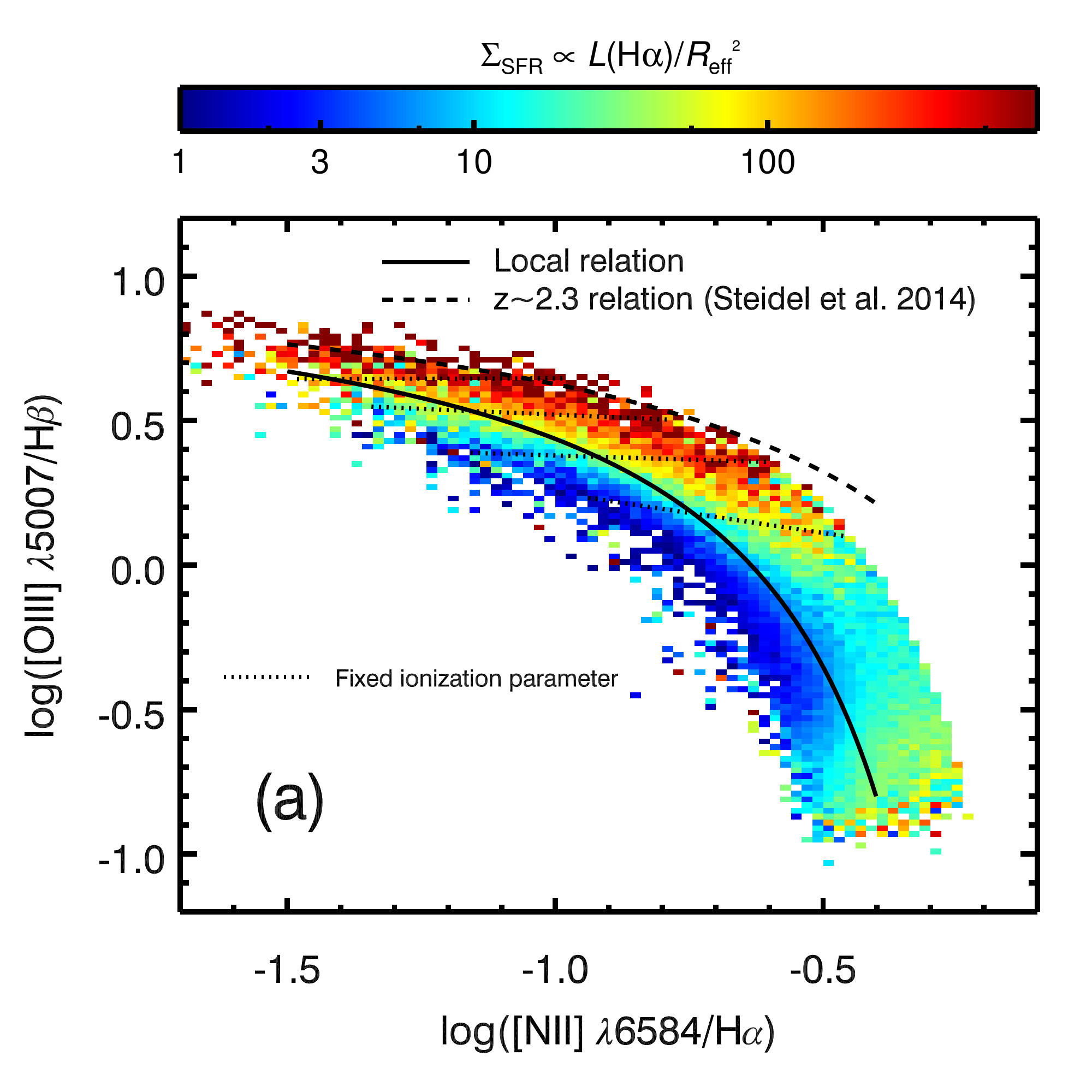} &
    \includegraphics[width=.45\textwidth]{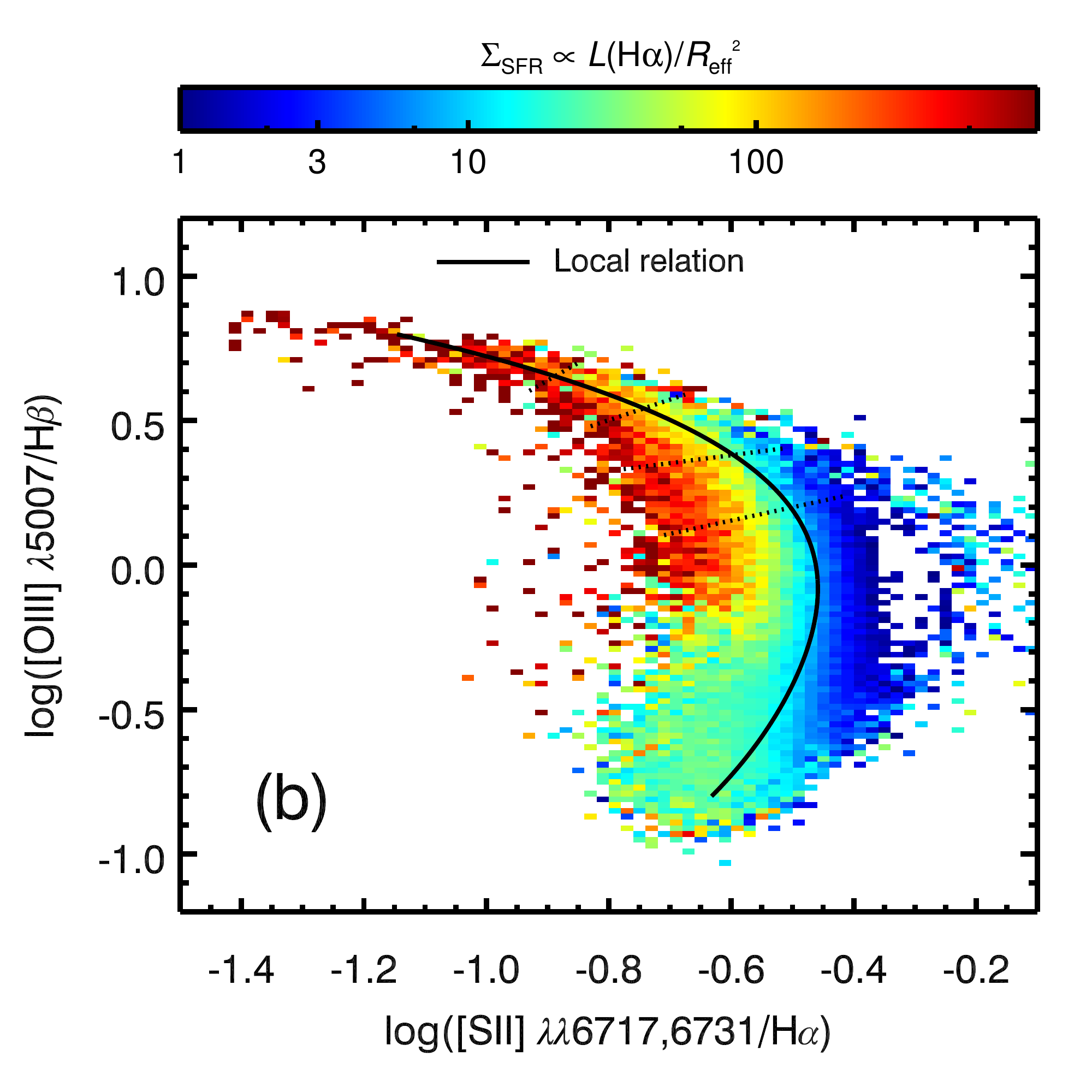} \\
       \includegraphics[width=.45\textwidth]{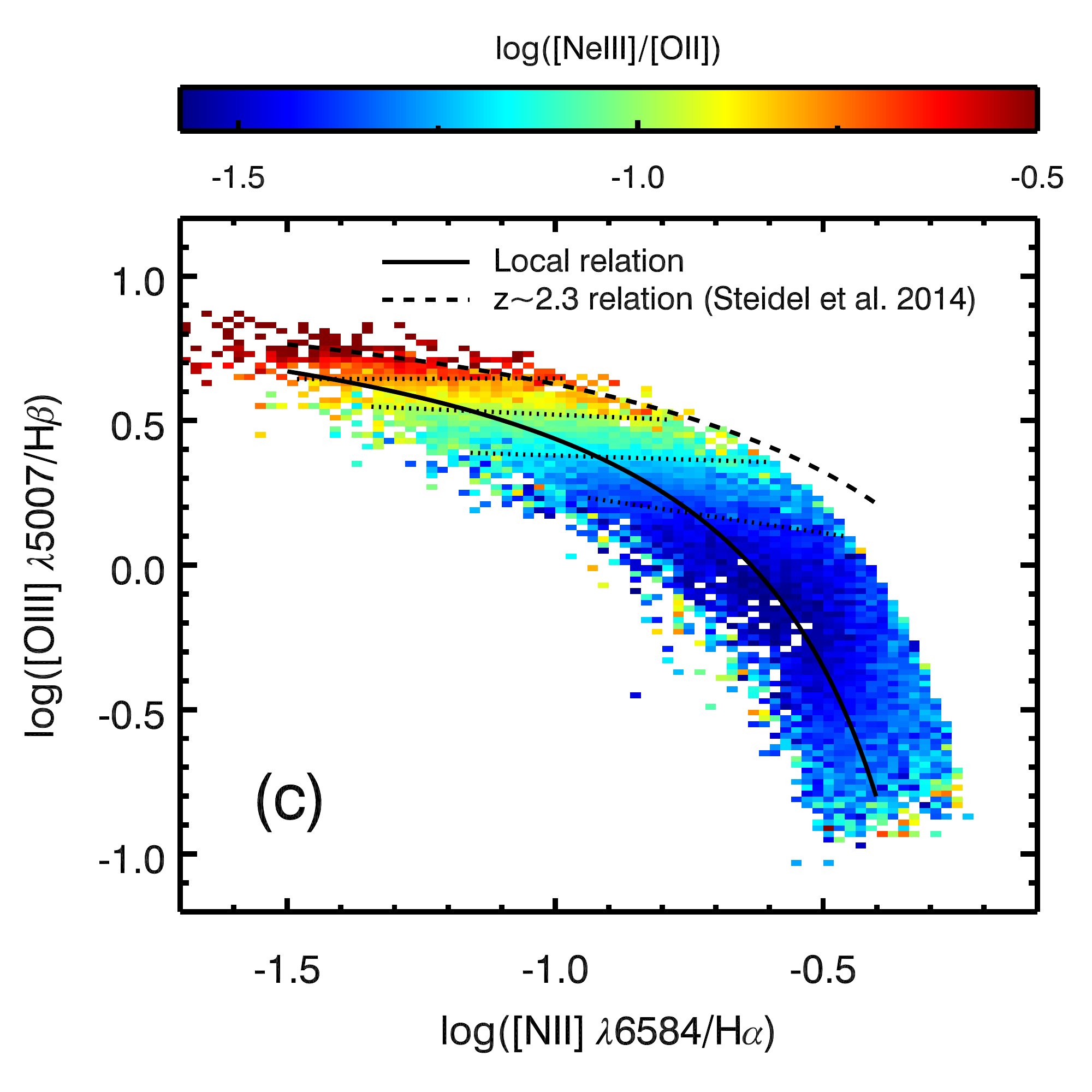} &
    \includegraphics[width=.45\textwidth]{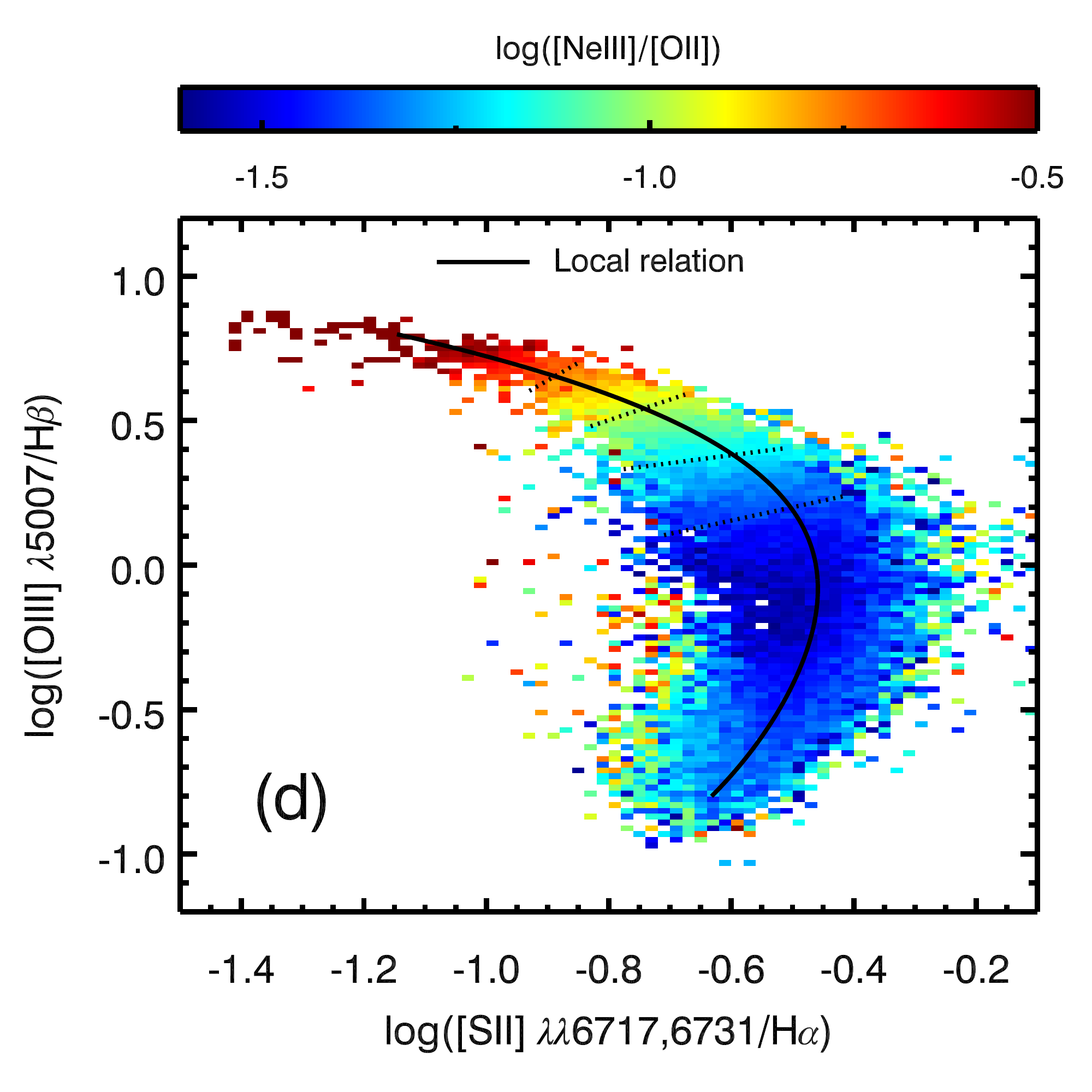} \\
  \end{tabular}
  \label{figure:sfrd}
  \caption{Using a sample of 104,256 star-forming galaxies from SDSS DR12, we show the median values of $L$(\ha)/\reff\ and [\neiii]$\lambda$3869/[\oii]$\lambda$3727 as a function of position in the O3N2 and O3S2 diagrams.  $L$(\ha)/\reff\ is an observable proxy for the surface density of star formation, \sigsfr, while [\neiii]/[\oii] is an observable proxy for ionization parameter \citep{Levesque14}. The median value of galaxies falling in pixels of size 0.02$\times$0.02 in each diagram is shown in color. The dotted lines indicate particular fixed ionization parameter levels based on the observed [\neiii]/[\oii] ratio. A normalization factor is applied to $L$(\ha)/\reff\ such that the minimum value shown is set to 1. \textbf{(a)} The trend in the O3N2 diagram is essentially the same as that uncovered by \citet{Brinchmann08b}. We see a strong segregation of galaxies depending on \sigsfr, such that high \sigsfr\ galaxies fall closer to the high-redshift locus.  \textbf{(b)} Intriguingly, we see a strong correlation between \sigsfr\ and position on the O3S2 diagram as well; however, the most strongly star-forming sources cluster on the \emph{left} side of the distribution in the O3S2 diagram, in contrast to the O3N2 diagram.  Also of note is the orthogonality between the lines of constant ionization parameter and the variation in \sigsfr, which we interpret as evidence that \sigsfr\ is not (directly) linked to ionization parameter.  \textbf{(c)} Log([NeIII]/[OII]) as function of position on the O3N2 diagram. The variation is primarily along the $y$ axis, confirming that the [\oiii]/\hb\ ratio is a relatively clean measure of ionization parameter.  \textbf{(d)} The variation of [\neiii]/[\oii] across the O3S2 diagram. Note that the variation in ionization parameter along the star-forming sequence in this diagnostic is essentially orthogonal to the variation observed with \sigsfr\ in panel (b).}
\end{figure*}

\begin{figure*}[htb]
\centering
  \begin{tabular}{@{}cc@{}}
    \includegraphics[width=.45\textwidth]{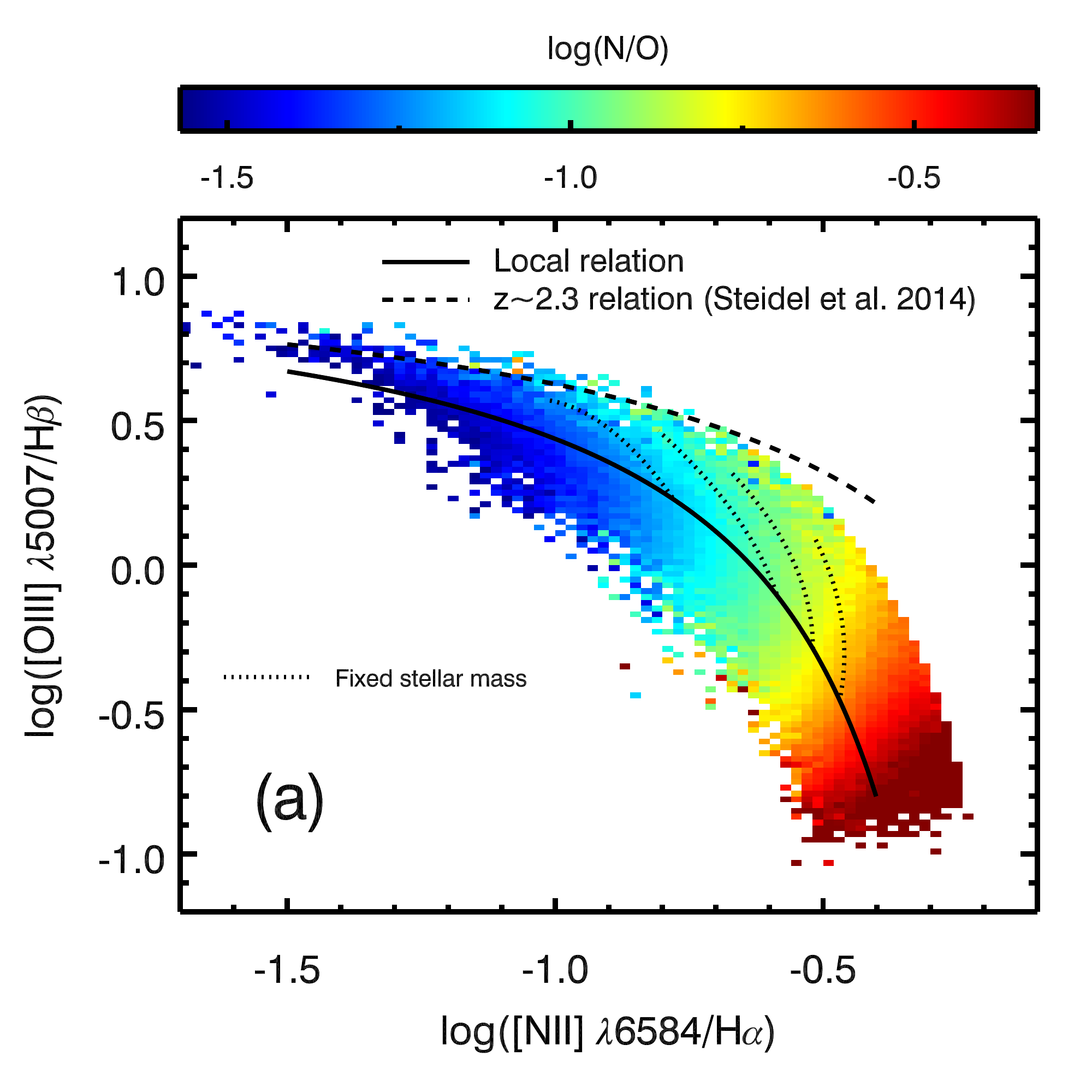} &
    \includegraphics[width=.45\textwidth]{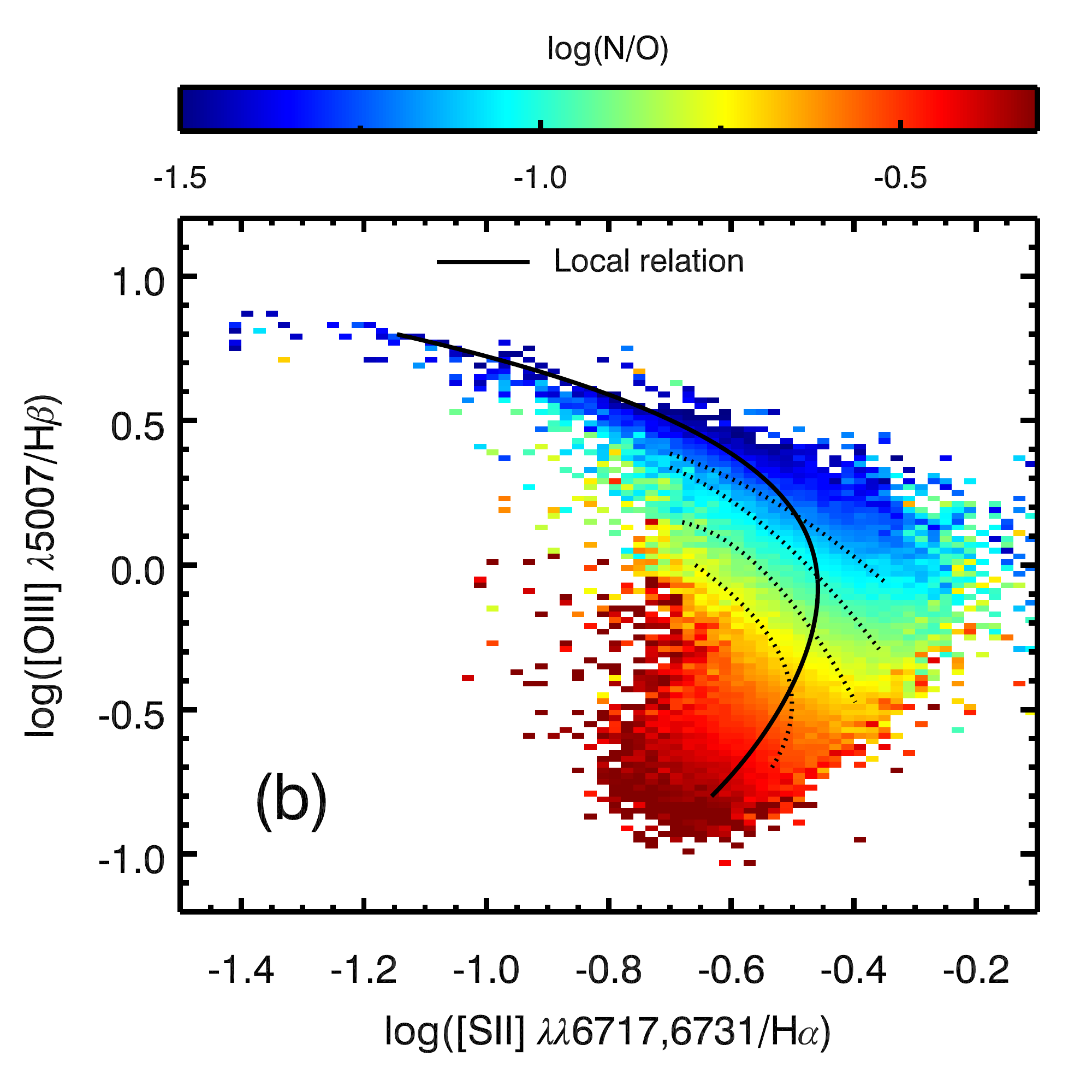} \\
        \includegraphics[width=.45\textwidth]{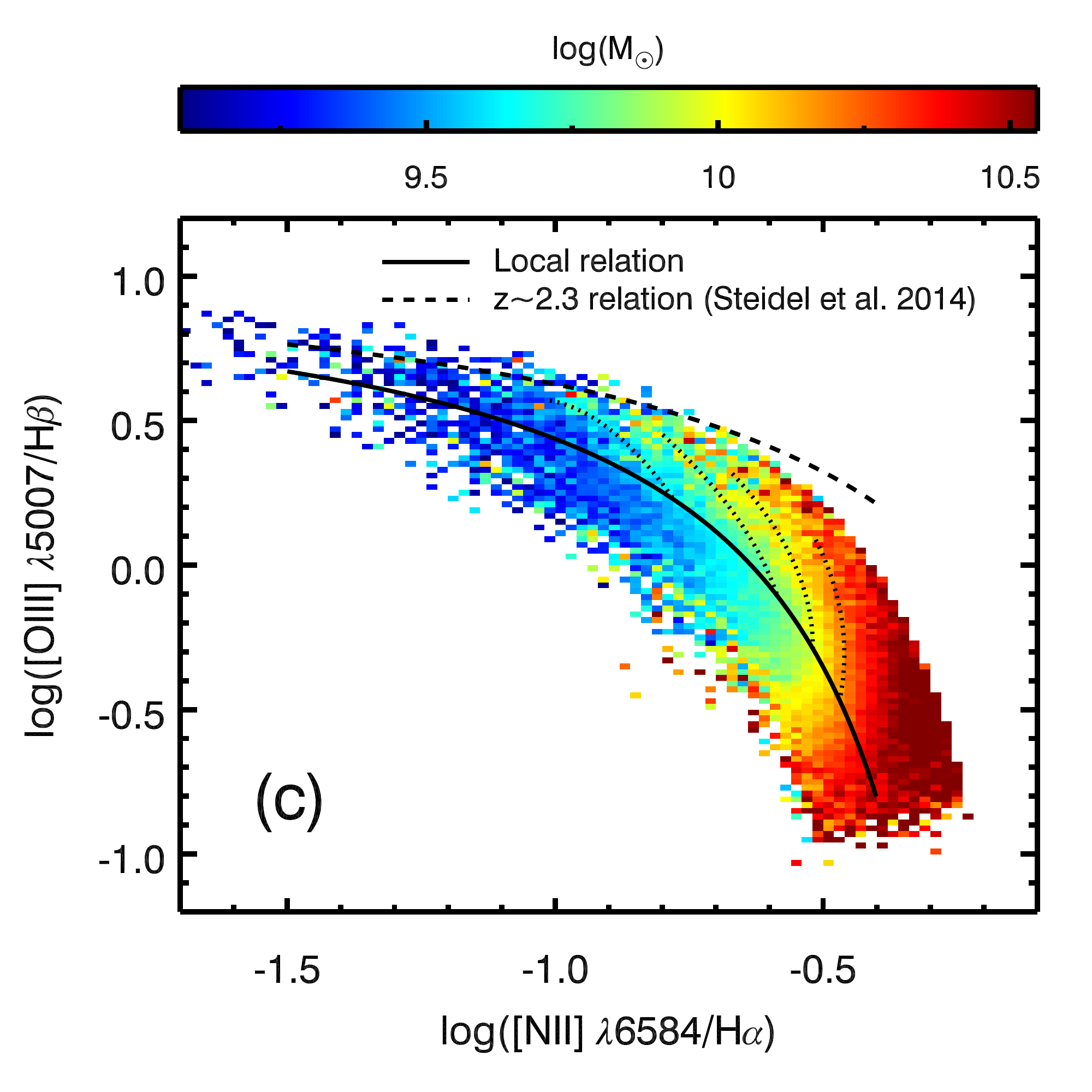} &
    \includegraphics[width=.45\textwidth]{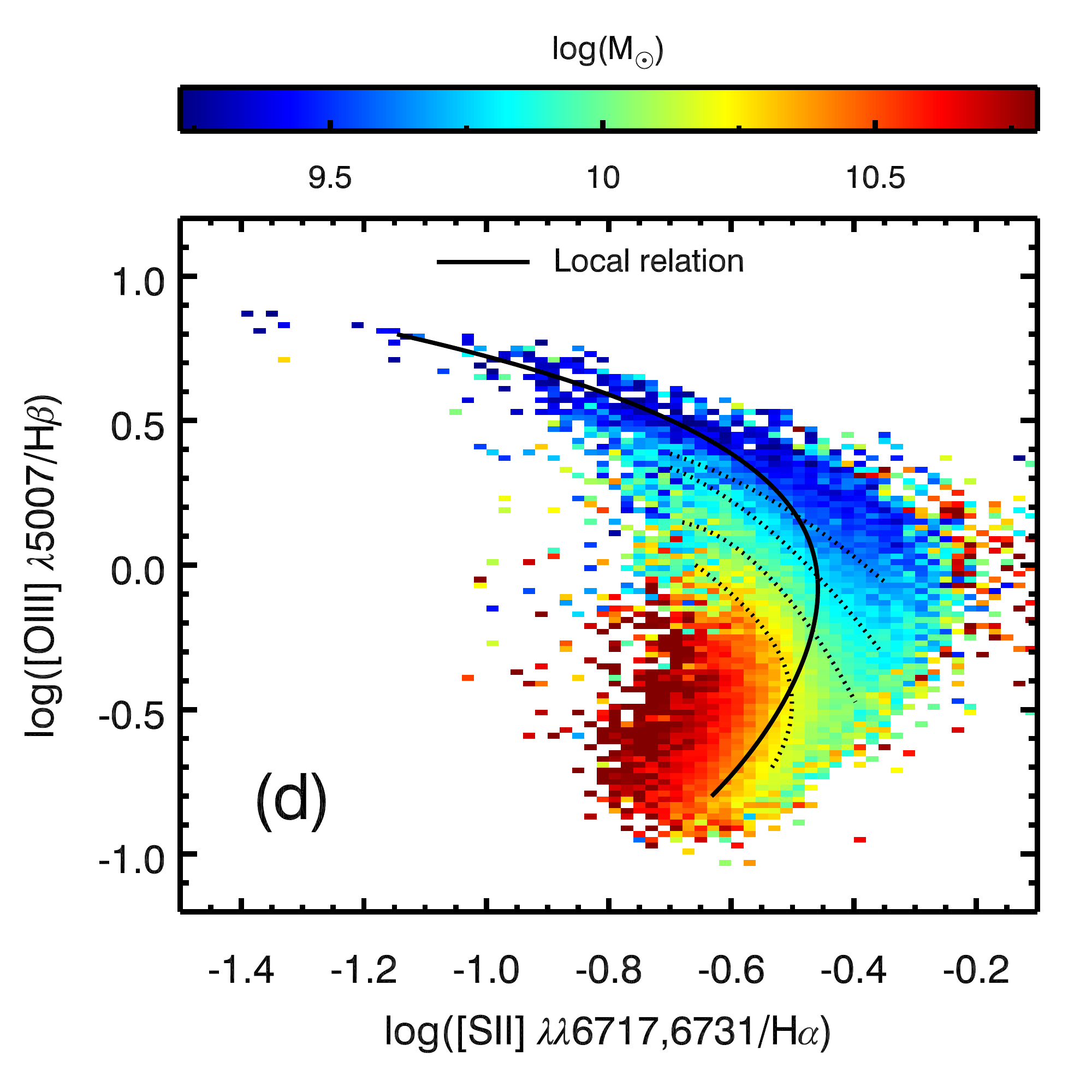} \\
  \end{tabular}
  \label{figure:no}
\caption{The median log(N/O) ratio and galaxy stellar mass as a function of position in the O3N2 and O3S2 diagnostic diagrams. The N/O ratio is computed using the dust-corrected ratio log(N$^{+}$/O$^{+}$) as in \citet{Thurston96}, assuming a nebular temperature of $10^{4}$~K. \textbf{(a)} The distribution of log(N/O) across the O3N2 diagram. For most [\oiii]/\hb\ values, moving to higher [\nii]/\ha\ also means moving to a higher N/O ratio; this illustrates how N/O can drive galaxy position on the $x$ axis of O3N2.  \textbf{(b)} A qualitative difference is apparent between the distribution of log(N/O) in O3S2 in comparison with O3N2. Generally, a lower log(N/O) correlates with being higher on the diagram, but there is a tilt to the correlation. \textbf{(c)} The median stellar mass of galaxies as a function of position in the O3N2 diagram. Note in particular the broadly similar distribution of stellar mass and N/O in panels (c) and (a), particularly in the star-forming wing of the distribution. On all panels, dotted lines indicate fixed mass values of log($M_{*}$) = 9.6, 9.8, 10.0, and 10.2~$M_{\odot}$, from top to bottom, respectively. Given that the correlation persists over a large range of galaxy stellar masses, ionization parameters, and star formation rates, it is strongly suggestive of a correlation between N/O and stellar mass more fundamental than the link between N/O and O/H, which is known to have a secondary dependence on SFR (e.g., \citealp{Brown16}). \textbf{(d)} The stellar mass distribution in O3S2 space. Again we see broad similarity between this distribution and log(N/O) in panel (b).}
\end{figure*}

\section{Empirical correlations in the diagnostic diagrams}

\subsection{Spectroscopic sample}
Our galaxy sample is drawn from the Sloan Digital Sky Survey (SDSS, \citealp{York00}) DR12 release \citep{Alam15}. We select galaxies with S/N~$>$~5 detections of the important emission lines [\oii]$\lambda$3726, [\oii]$\lambda$3729, \hb,  [\oiii]$\lambda$5007, \ha, [\nii]$\lambda$6584, [\sii]$\lambda$6717, and [\sii]$\lambda$6731. We also restrict our sample to galaxies classified as star-forming on the BPT diagram. This cut introduces an artificial upper boundary on galaxies in O3N2, but the underlying trends in the distribution are not affected. In order to minimize the effect of the finite fiber aperture (3$''$ for the SDSS spectra used), we restrict our sample to galaxies above $z=0.05$, such that the 3$''$ fiber covers at least the central $\sim$1.5 kpc of the galaxy. In total, our sample comprises 104,256 galaxies with high-quality spectra and emission line flux measurements from SDSS. 

\subsection{Physical properties considered}
We examine how four  physical properties correlate with
galaxy position on the O3N2 and O3S2 diagrams. It is helpful to examine the O3S2 diagram as well because it does not suffer from uncertainties arising from variations in the N/O abundance ratio (sulfur is an alpha element so its relative abundance closely tracks oxygen). The four physical properties we consider are:

\begin{enumerate}
\item{The surface star formation rate density, \sigsfr, as encoded by the \ha\ luminosity divided by the square of the effective radius
    ($L$(\ha)/\reff). This value is closely related to both specific star formation rate and EW(\ha).}
\item{The ionization parameter $U$, using the line ratio [\neiii]$\lambda$3869/[\oii]$\lambda\lambda$3726,3729 as an observable proxy  \citep{Levesque14}.
    This ratio is insensitive to
    metallicity variation, as the neon abundance is expected to closely track oxygen. Moreover, the lines are close in wavelength, reducing the impact of an uncertain reddening correction, and it avoids the use of the [\oiii]$\lambda$5007 line, which is already in the diagnostic diagrams.}
\item{The N/O ratio, as determined from the dust-corrected ratio
    log(N$^{+}$/O$^{+}$) as in \citet{Thurston96}, assuming a nebular temperature of $10^{4}$~K.}
\item{The stellar masses provided by SDSS in the MPA-JHU spectroscopic catalog \citep{Kauffmann03}, which were derived assuming a \citet{Kroupa01} initial mass function (IMF).}
\end{enumerate}

We chose these quantities because they are all (nearly) directly measurable, and they are likely to be linked to the proximate causes of galaxy emission line ratios, thus yielding insight into the origin of 
the sequence and the cause of its variation with redshift. The ionization parameter (roughly a measure of the relative number of ionizing photons to hydrogen atoms) has been discussed extensively in a number of papers; for an excellent overview we refer the reader to \citet{Sanders16}. We have chosen not to discuss electron density in the following analysis, but we note that it has a weak correlation with position on the diagrams that tracks \sigsfr, as also found by \citet{Brinchmann08b}.

For all quantities computed using emission lines fluxes, we first apply a reddening correction using the measured Balmer decrement \ha/\hb\ in conjunction with a Calzetti extinction curve \citep{Calzetti00}. In the BPT diagrams we compute the median value of each of the physical properties for galaxies in pixels of size 0.02$\times$0.02. In \S2.3-\S2.7 we describe the observed correlations and discuss some implications, and in \S3 we provide a more detailed interpretation relating the local sample to what is observed at high redshift.

\subsection{Dependence of the diagnostic ratios on \sigsfr}
Panels (a) and (b) in Figure~1 illustrate the striking segregation of galaxies in both the O3N2 and O3S2 diagnostic diagrams according to their value of $L$(\ha)/\reff. Local galaxies with high values of \sigsfr\ tend to fall closer to the high-redshift locus of star-forming galaxies, which was determined by \citet{Steidel14} from MOSFIRE spectra from the KBSS survey. \citet{Brinchmann08b} uncovered essentially the same trend in the O3N2 diagram, concluding that an enhanced ionization parameter in strongly star-forming galaxies was likely behind the effect.

The behavior of galaxies in the O3S2 diagram is somewhat surprising. Here a clear segregation is also seen, but
in a different sense, with the most strongly star-forming galaxies
falling on the left side of the star-forming sequence (lower [\sii]/\ha\ at fixed [\oiii]/\hb). A similar trend was recently noted by \citet{Kashino16} for $z\sim1.6$ galaxies in the FMOS-COSMOS survey, suggesting that this behavior appears in the high redshift galaxy samples as well (particularly for galaxies with masses $M_{*}>10^{10.3}M_{\odot}$). Kashino~et~al. attribute this trend to a higher ionization parameter, which could increase the excitation of S$^{+}\rightarrow\mathrm{S}^{++}$ and drive down the [\sii]/\ha\ ratio. However, the empirical correlaton with ionization parameter discussed in the next section challenges this interpretation.

\begin{figure*}[htb]
\centering
    \includegraphics[width=.81\textwidth]{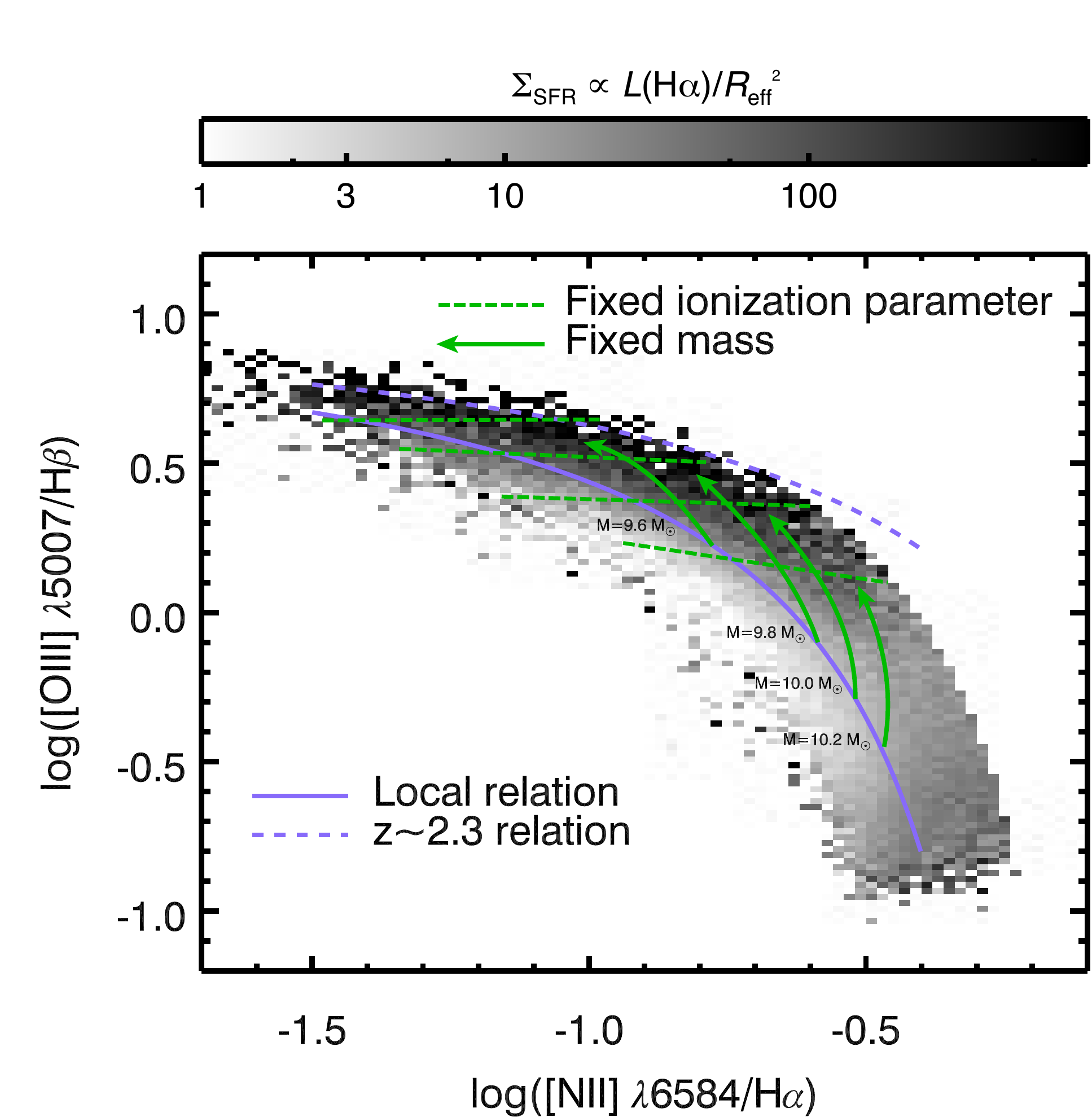} 
  \label{figure:ssfr_with_lines}
  \caption{A view of how multiple galaxy parameters vary simultaneously in O3N2 space. Paths of constant stellar mass are shown originating on the local star-forming locus, with arrowheads pointing toward the high-redshift locus. At a fixed stellar mass, both \sigsfr\ and ionization parameter increase to drive the galaxies toward the high-redshift locus. As illustrated in the next plot, the explanation for why these enhanced \sigsfr\ galaxies move off of the local star-forming abundance sequence is that their N/O ratios are higher than typical local galaxies at the same [\oiii]/\hb\ ratio.}
\end{figure*}

\begin{figure*}[htb]
\centering
    \includegraphics[width=.81\textwidth]{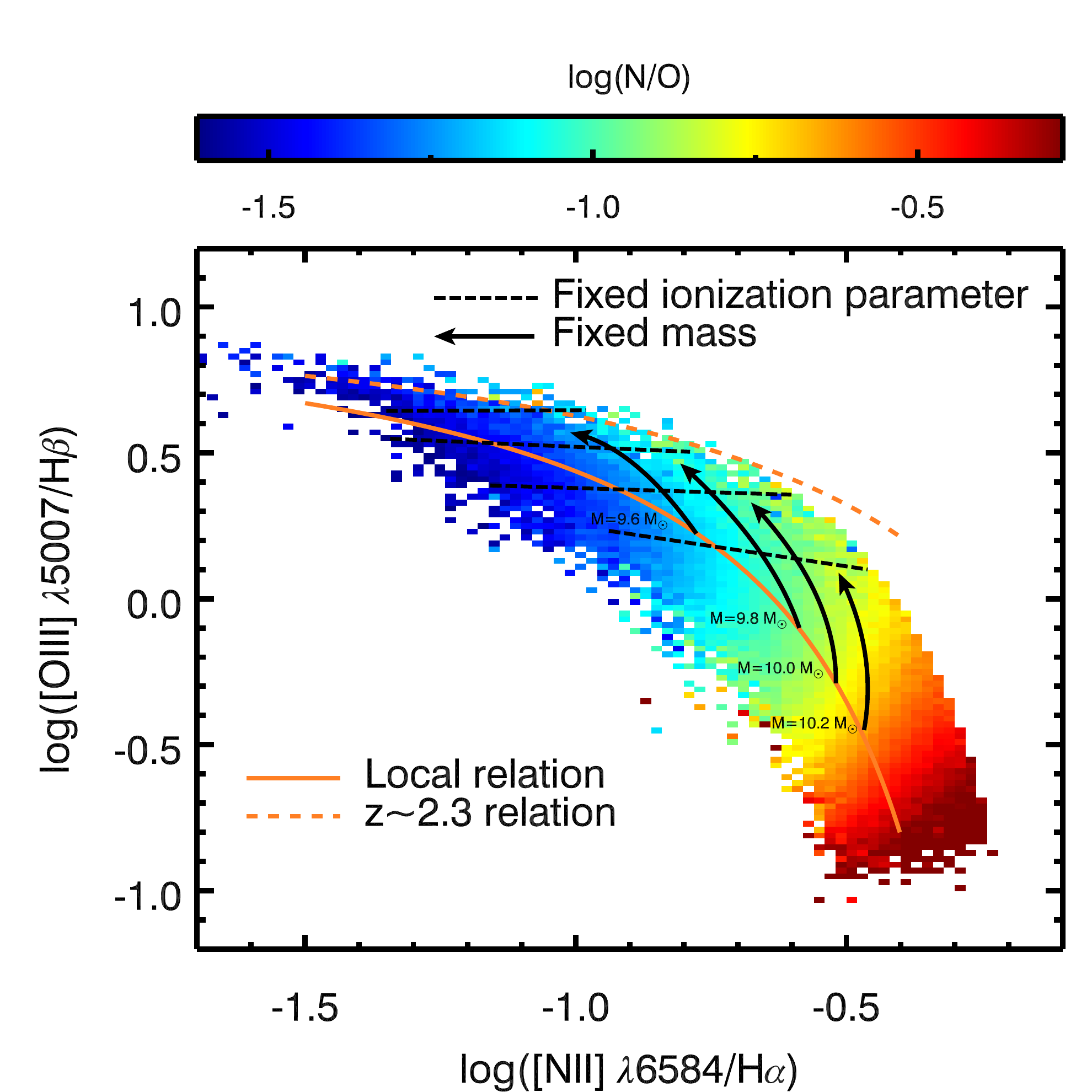} 
  \label{figure:no_with_lines}
  \caption{Another view of how multiple parameters vary simultaneously in O3N2 space, this time with the N/O ratio as the backdrop. Note that galaxies of a fixed stellar mass remain at an essentially constant N/O as they move off of the local locus toward the high redshift locus. The higher N/O values at a fixed [\oiii]/\hb\ cause them to fall to the right of the local relation, and closer to the high redshift relation. At fixed ionization parameter, moving toward the high-redshift locus is associated with having both higher N/O and higher stellar mass.}
\end{figure*}

\subsection{Dependence on ionization parameter}
Panels (c) and (d) in Figure~1 illustrate how ionization parameter $U$ (as traced by [\neiii]/[\oii]) correlates with the line ratios. In both diagrams,  this ratio is primarily predictive of the position along the $y$~axis, confirming that [\oiii]/\hb\ is closely linked to ionization parameter. 

Importantly, the variation of ionization parameter, particularly in the O3S2 diagram, is essentially \emph{orthogonal} to the observed segregation with \sigsfr. This is highlighted in panels (a) and (b) of Figure~1, where dotted lines indicate fixed ionization parameter levels as determined from panels (c) and (d). As can be seen, the same ionization parameter is found for a large range of \sigsfr\ values. If ionization parameter and \sigsfr\ were closely linked, as has been hypothesized based on observations of high redshift galaxies and their local analogs, we would expect to see a trend in [\neiii]/[\oii] that mimics the trend with \sigsfr; the fact that we see essentially no relation favors the idea that ionization parameter is not (directly) dependent on the star formation intensity. 

\subsection{Dependence on N/O}

Panels (a) and (b) of Figure~2 illustrate the equally strong trends in the diagnostic diagrams with the nitrogen-to-oxygen ratio, N/O. This correlation is not unexpected, as N/O is known to track the oxygen abundance. Moreover, the N/O ratio directly impacts where a galaxy falls on $x$~axis of the O3N2 diagram. However, the details of the correlation are quite revealing. First we note the qualitative difference in the variation of N/O across the ionization sequence in the O3N2 diagram relative to the O3S2 diagram. This is largely due to the interplay between the N/O ratio and the position of a galaxy on the $x$~axis of O3N2. 

At a fixed [\oiii]/\hb, the N/O ratio increases as galaxies move to the right in O3N2, toward the high-redshift locus of star-forming galaxies. This trend holds over almost the entire star-forming sequence. The straightforward implication is that, at a fixed [\oiii]/\hb\ (and hence ionization parameter), local analogs to high redshift galaxies approach the high redshift locus because they have higher N/O ratios. In contrast, the N/O ratio does not vary as much as galaxies move up in [\oiii]/\hb\ at a fixed [\nii]/\ha.

The behavior of N/O in the O3S2 diagram (panel (b) of Figure~2) shows an upward trend with decreasing N/O, as expected from the correlation of N/O with metallicity and the anti-correlation of metallicity with ionization parameter. Interestingly, there is a pronounced tilt of the correlation across O3S2. This tilt implies that the same N/O can correspond to different ionization parameters, which was also evident from the joint behavior of N/O and ionization parameter in O3N2.

\subsection{Dependence on stellar mass}
Panels (c) and (d) of Figure~2 show a clear trend with galaxy stellar mass in both O3N2 and O3S2. Notably, the stellar mass trend closely mirrors the trend with N/O over much of the diagrams, as highlighted with dotted lines of fixed mass on panels (a) and (b) as derived from panels (c) and (d). The divergence of the two distributions in the bottom part of the diagrams is likely due to the known flattening of the N/O-stellar mass relation at high masses \citep{Perez09}. The joint correlation of N/O and stellar mass may be thought to reflect the mass-metallicity relation, as N/O is known to be metallicity-dependent (at least for metallicities above $\sim$0.2~Z$_{\odot}$, \citealp{Henry00}). However, while N/O generally tracks O/H, it is now known to show a strong secondary dependence on SFR (e.g., \citealp{Andrews13, Brown16}). The similarity of the correlations of stellar mass and N/O across much of the diagnostic diagrams in Figure~2 indicates that these two quantities trace each other closely over a range of ionization states, SFRs, and metallicities, with very little secondary dependence on these other quantities.

\subsection{Summary of key observations}

Here we summarize key observations to be made based the empirical correlations in Figures~1 and 2: 

\begin{itemize}
\item{\sigsfr\ is strongly correlated with position on both diagrams. Galaxies with very high \sigsfr\ fall near the high redshift locus in O3N2, while they fall on the lower left side of the O3S2 diagram.}
\item{Ionization parameter (as measured by [\neiii]/[\oii]) is closely correlated with [\oiii]/\hb\ (the $y$ axis of both diagrams) and shows little or no (direct) dependence on \sigsfr.}
\item{At a fixed [\oiii]/\hb, moving to the right on the O3N2 diagram means moving to increasing stellar mass, \sigsfr, and N/O ratio at a roughly constant ionizaton parameter.}
\item{At a fixed [\nii]/\ha, moving up on the O3N2 diagram means moving to increasing ionization parameter and \sigsfr\ at a roughly constant stellar mass and N/O ratio.}
\item{The N/O ratio and the stellar mass closely track each other over most of the diagnostic diagrams.}
\end{itemize}

These observations are significant clues as to the nature of the BPT shift in O3N2, which we now explore in more detail.

\section{Physical interpretation}

In the preceding section we used purely empirical correlations in the SDSS data to highlight significant trends in the O3N2 and O3S2 diagrams. Here we attempt to understand the redshift evolution of emission line ratios in the O3N2 diagram (we defer a discussion the intriguing behavior in O3S2 to \S5.3). Our working assumption is that the physical factors that push local galaxies toward the high-redshift part of the O3N2 diagram are the same as those driving the observed position of high redshift galaxies.

\subsection{Dissecting the O3N2 diagram}

In \S2 we showed a number of purely empirical correlations of different physical quantities with position on the O3N2 diagnostic diagram, which together represent a complex, multidimensional set of relationships. Here we attempt to interpret the relationships by examining the O3N2 diagram in greater detail.

In Figure~3 we show how ionization parameter, stellar mass, and \sigsfr\ interrelate. The same lines of constant ionization parameter and stellar mass derived from Figures~1 and 2 are overlaid, with arrowheads on the lines of constant stellar mass indicating movement off the local sequence and toward the high-redshift locus. At a fixed stellar mass, movement toward the high-redshift sequence is associated with both higher \sigsfr\ and higher ionization parameter. We conclude that it is indeed the case that galaxies near the high-redshift locus have higher ionization parameters than typical galaxies \emph{of the same stellar mass}, and that this is linked to their higher \sigsfr\ values. Therefore, while no \emph{direct} correlation between \sigsfr\ and ionization parameter is observed, these quantities do in fact correlate with each other at a fixed stellar mass.

In Figure~4 we show how ionization parameter, stellar mass, and log(N/O) interrelate in O3N2. Note that galaxies at a fixed stellar mass continue to have roughly the same log(N/O) value as they move off the local relation and towards the high-redshift locus. We conclude from this plot that galaxies near the high-redshift locus have both higher stellar masses and higher log(N/O) values \emph{at a fixed ionization parameter} than galaxies falling on the local star-forming sequence. The implication is that the higher N/O values are what drive these galaxies to be displaced to the right on the [\nii]/\ha\ axis in comparison with lower-mass and lower N/O galaxies with the same ionization parameter. 

\begin{figure}[htb]
\centering
    \includegraphics[width=.45\textwidth]{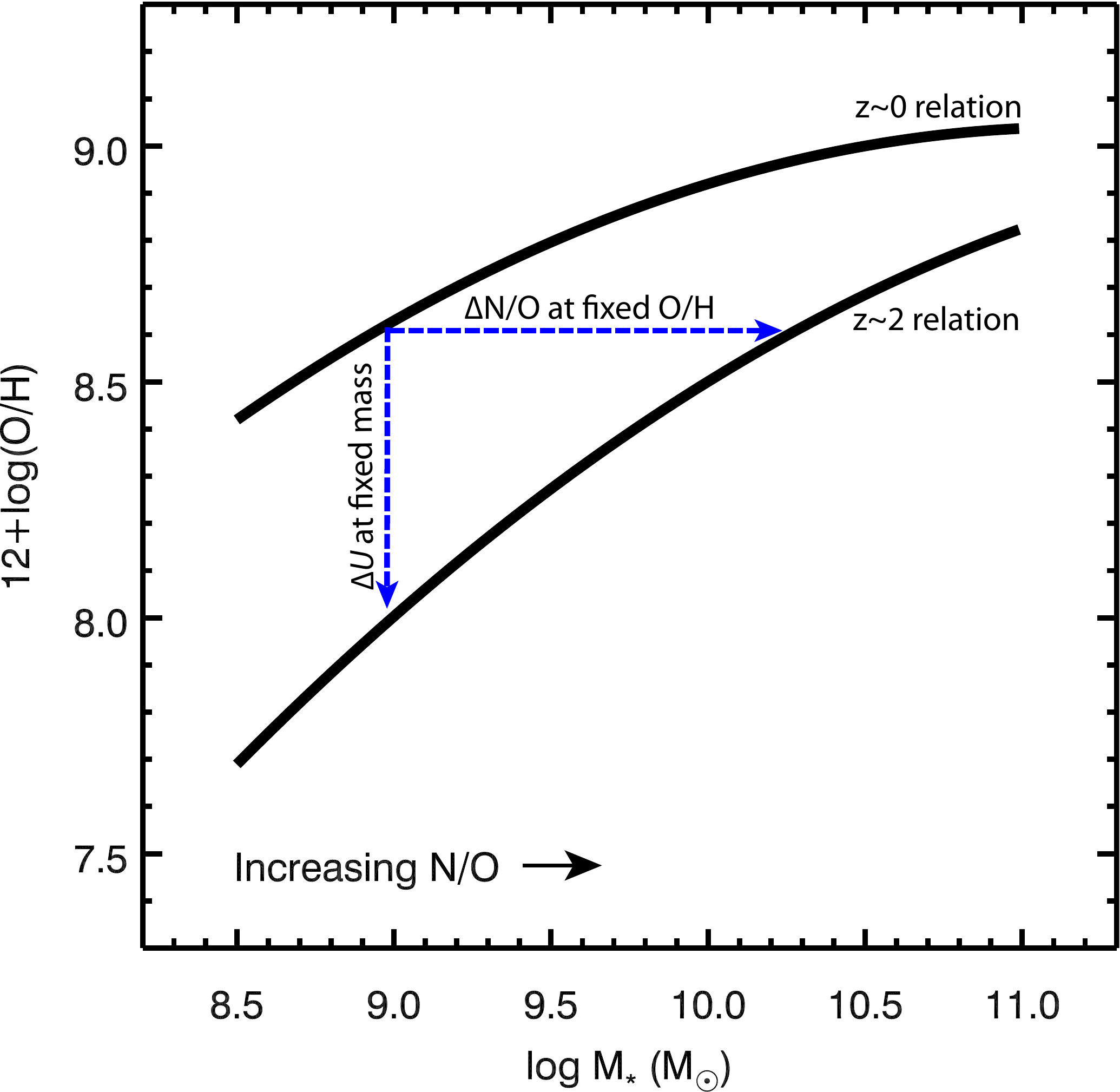} 
  \label{figure:mz_no}
  \caption{Schematic showing how the evolving mass-metallicity (MZ) relation, together with a fixed (or slowly varying) relation between the N/O ratio and stellar mass, produces an apparent nitrogen enhancement in high-redshift galaxies at a fixed oxygen abundance. The evolving MZ relation also causes a changing ionization parameter $U$ at fixed mass, assuming $U$ and metallicity are linked. The parameterizations of the MZ relation are from \citet{Maiolino08}.}
\end{figure}

\subsection{Origin of the BPT offset: A tight relation of N/O with stellar mass}

We postulate that the N/O ratio of galaxies depends principally on total stellar mass, rather than on the gas-phase metallicity, while the ionization parameter is driven almost \emph{entirely} by gas-phase metallicity. The N/O-$M_{*}$ link is supported by the similar distribution of these quantities across much of the diagnostic diagrams, indicating little or no secondary dependence on metallicity, SFR, or ionization state. In the next section we will discuss physical reasons why N/O might be expected to be more fundamentally tied to stellar mass than to metallicity.

The idea that ionization parameter is driven primarily by gas phase metallicity is supported by the observation that \emph{U} and \sigsfr\ show essentially no direct correlation. Ionization parameter depends on the rate of production of ionizing photons $Q_{0}$, the electron density $n_{e}$, and the gas filling factor $\epsilon$ as follows: \begin{equation} U\propto Q_{0}^{1/3}n_{e}^{1/3} \epsilon^{2/3} \end{equation} Both $n_{e}$ and $\epsilon$ could be expected to correlate with \sigsfr\ ($n_{e}$ is in fact known to show such a correlation, although it is rather weak \citep{Brinchmann08b}; we verified the weak correlation with our data as well). However, given the apparent lack of a significant direct correlation between \emph{U} and \sigsfr, we argue that density and geometric effects are secondary, and the primary factor that changes the ionization parameter is $Q_{0}$, which is most affected by the metallicity of the ionizing O and B stars \citep{Dopita06, Kewley13}. \citet{Sanders16} recently arrived at the same conclusion from an extensive analysis of data from the MOSDEF survey. 

The following scenario can explain the offset of local high-redshift analogs in the O3N2 diagram. Local galaxies with significantly enhanced SFR values have experienced a recent infall of pristine gas, lowering their gas phase O/H and increasing their ionization parameters. The infall of pristine hydrogen gas does not, however, affect their N/O ratios, so that, while they move upward on the O3N2 diagram, they are offset toward higher values of [\nii]/\ha\ than ``typical" galaxies of the same [\oiii]/\hb.

If the N/O-$M_{*}$ relation is more fundamental than N/O-O/H, we also expect it to be more nearly invariant with redshift (although we will discuss evidence that suggests a modest redshift evolution of N/O-$M_{*}$ in the next section). Therefore, given the rapid evolution of the mass-metallicity relation with redshift (e.g., \citealp{Erb06, Maiolino08, Zahid13, Henry13, Ly16}), we would expect N/O to display a different scaling with O/H as a function of redshift, in precisely the way that is observed (Figure 5).  At a fixed O/H, a galaxy will have a higher mass at high redshift, and thus a higher N/O than a galaxy with the same O/H in the local universe. 

Two equivalent ways of explaining the shift of high-redshift galaxies in the O3N2 diagram are therefore as follows:

\begin{enumerate}
\item{At a fixed stellar mass, a high redshift galaxy has a higher SFR, lower O/H, and higher ionization parameter than local galaxies, pushing it upward on the O3N2 diagram. There is an offset from the local star-forming sequence because the galaxy has a higher mass -- and therefore N/O ratio -- than typical local star-forming galaxies at the same [\oiii]/\hb\ level.}
\item{At a fixed gas-phase metallicity/ionization parameter, a high redshift galaxy will have a higher stellar mass and N/O ratio than low-redshift galaxies, causing it to be displaced to the right on the O3N2 diagram.}
\end{enumerate}

Ultimately, both enhanced ionization parameters (at a given stellar mass) and higher N/O values at a fixed O/H are required to explain the observed O3N2 shift with redshift, but these effects should be understood in the context of a nearly invariant N/O--$M_{*}$ relation coupled with an evolving mass-metallicity relation.

\begin{figure*}[htb]
\centering
    \includegraphics[width=.77\textwidth]{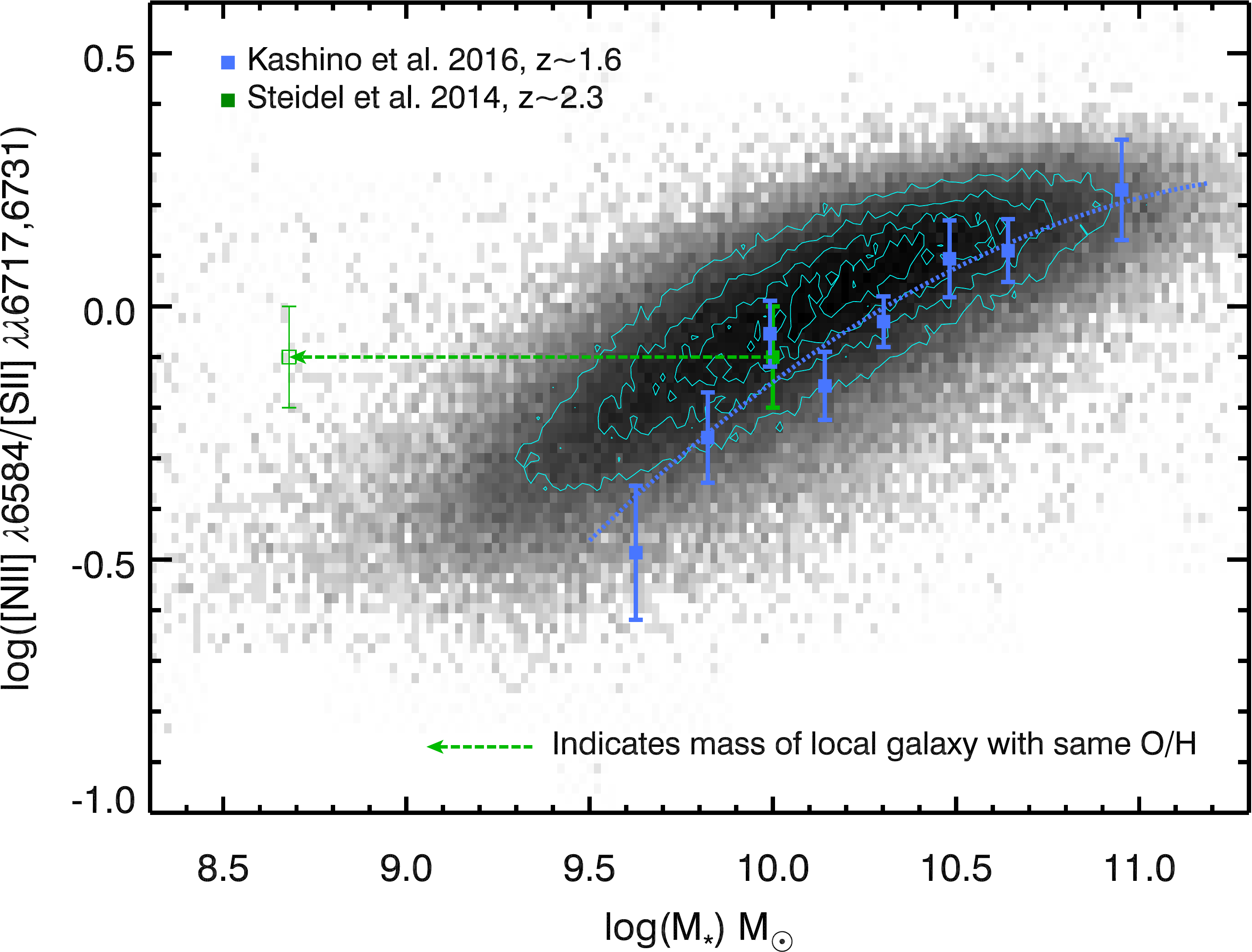} 
  \label{figure:ns_mass}
  \caption{We plot the local relation of [\nii]$\lambda$6584/[\sii]$\lambda\lambda$6717,6731 (N/S -- considered to be a good proxy for N/O) with stellar mass, in comparison with the high-redshift relation from \citet{Steidel14} (from average reported values for the KBSS sample) and \citet{Kashino16} from stacked spectra from the FMOS-COSMOS survey at $z\sim1.6$. The FMOS-COSMOS masses have been converted from a \citet{Salpeter55} IMF to \citet{Kroupa01} IMF for comparison with the SDSS masses. The dotted blue line is a fit to all of the high-redshift values. A shift of $\sim$0.1 dex of the high-redshift samples relative to local galaxies is apparent. At the masses sampled the high-redshift galaxies are almost all on the secondary N/O production regime, despite the low metallicities of these galaxies at high redshift. The arrow indicates the mass of a local galaxy expected to have the same O/H as the $z\sim2.3$ point shown, using the parameterization of the mass-metallicity relation from \citet{Maiolino08}. Clearly, a local galaxy with the same N/O would be considered to be significantly elevated in N/O versus O/H; in other words, while the N/O of the points of the high-redshift galaxies are slightly lower than local galaxies at a fixed mass, they are actually at quite elevated N/O values \emph{for their O/H values}. While the N/O-$M_{*}$ relation evolves modestly with redshift, the N/O-O/H relation varies strongly. The origin of the small shift of the N/O-$M_{*}$ relation with redshift is plausibly related to galaxy age, as discussed in the text.}
\end{figure*}

\section{Is a tight relation between N/O and stellar mass reasonable?}

There is a rich body of literature on the nucleosynthetic origin of nitrogen and the N/O versus O/H relation  (e.g., \citealp{Edmunds78, Vila93, Kobulnicky98, Henry00, Pilyugin03, Koppen05, vanZee06, Perez13}).  Nitrogen is thought to have both a primary and a secondary origin. Primary nitrogen is mostly produced in CNO burning from the carbon/oxygen generated by the star itself during its helium burning phase, while secondary nitrogen is generated through CNO burning of pre-existing carbon and oxygen that were in the star at its formation. The secondary production of nitrogen is the reason for the observed scaling of the N/O ratio with metallicity. Studies in the local universe indicate that nitrogen is a primary
    element at low metallicity (12+log(O/H)~$\lesssim$~1/3~$Z_{\odot}$) and scales linearly with metallicity above that value \citep{vanZee98}.

While nitrogen is known to be produced in the CNO cycle burning that occurs in intermediate and massive stars, the chief uncertainty in the origin of nitrogen has to do with which stars most effectively return their nucleosynthetic products to the ISM \citep{Henry00}. Significant time lags between star formation and the return of nitrogen to the ISM can also exist, depending on the stars most responsible for nitrogen enrichment. For example, if intermediate mass stars ($\sim$1-8~$ M_{\odot}$) are the main sources of nitrogen, there will be a delay of hundreds of millions of years between a starburst and the return of nitrogen to the ISM, while massive stars would enrich the ISM almost instantaneously by comparison.

Intuitively, the N/O ratio should have more of a ``closed-box" character than O/H, since it is insensitive to inflows of pristine gas as well as outflows of metals (assuming there is no significant differential expulsion of N and O). The instantaneous gas-phase metallicity of galaxies, on the contrary, depends strongly on the combined effects of pristine gas inflows coupled with the expulsion of metals through outflows. Therefore,  the N/O-O/H relation can show large variations, particularly from substantial gas inflows \citep{Koppen05}.  Assuming that the evolution of the mass-metallicity relation (MZR) of galaxies with redshift to some extent reflects increasing gas fractions due to rapid infall of fresh hydrogen, a redshift-invariant N/O-O/H relation would be surprising. On the other hand, the stellar mass of a galaxy is a rough measure of the total amount of stellar processing that has occurred. It stands to reason that the N/O ratio should be more linked to this quantity than to O/H.

\subsection{Observational evidence}
Observational evidence suggesting N/O is more fundamentally tied to mass than to gas-phase metallicity exists. For example, there is an observed secondary dependence of the N/O-O/H relation on SFR \citep{Andrews13, Brown16}. This secondary dependence on SFR is understandable if SFR increases and O/H decreases in response to the inflow of pristine gas. On the other hand, the N/O-$M_{*}$ relation shows essentially no secondary dependence on SFR \citep{Perez13}, and is generally tighter than the N/O-O/H relation (\citet{Andrews13}, Figure~14). Our analysis also shows that N/O and stellar mass are correlated over a broad range of other galaxy properties. The close link of N/O with stellar mass was also noted by \citet{Wu13}.

While the N/O-$M_{*}$ relation seems to be more fundamental than N/O-O/H, there is some evidence that it varies with redshift. \citet{Perez13} used the [\nii]/[\sii] ratio (N/S, a good proxy for N/O) to argue that the N/O-$M_{*}$ relation evolves modestly over the interval $z\sim0.1$ to $z\sim0.3$, in the manner expected if N/O tracks O/H. In addition, \citet{Kashino16} presented data based on spectral stacks suggesting that the N/S-$M_{*}$ relation evolves as a function of stellar mass between $z\sim0$ and $z\sim1.6$, with the strongest evolution at the low mass end (in fact, their data shows very little offset between the local and high-redshift relation at masses $\gtrsim10^{10.5}M_{\odot}$).  However, the stellar mass estimates were based on a \citet{Salpeter55} IMF, while the SDSS masses assume a \citet{Kroupa01} IMF. If the mass estimates are put on the same IMF scale as described in \citet{Speagle14}, the shift in the N/S-$M_{*}$ relation is significantly smaller. The average reported stellar mass (10$^{10} M_{\odot}$)  and N/S ratio ($-0.1\pm0.1$) of the KBSS sample at $z\sim2.3$ puts those galaxies close to the local relation as well. In Figure~6 we show the N/S distribution of local SDSS galaxies along with the FMOS-COSMOS and KBSS points. There is a slight downward shift (particularly at low masses) of the high-redshift N/S values at a fixed mass of $\sim$0.1 dex, which likely indicates some evolution of the N/O-$M_{*}$ relation with redshift. 

However, the relationship does not vary nearly as much as would be expected based on the evolution of the mass-metallicity relation. This is illustrated with an arrow in Figure~6 showing where a galaxy of similar \emph{metallicity} as the KBSS point would sit on the mass axis, using the MZR parameterization from \citet{Maiolino08}. While the N/S value of the KBSS point is slightly lower than the local N/S-$M_{*}$ relation, it is significantly elevated for its gas-phase metallicity. Indeed, a local galaxy of similar metallicity would be expected to be in the primary nitrogen regime, while the KBSS point and all but one of the FMOS-COSMOS points are clearly well into secondary nitrogen production.

Figure~7 illustrates in another way that the high-redshift samples will appear to have elevated N/O at fixed O/H. We show the median mass offsets computed between the local SDSS sample and the KBSS sample at eight different values of [\oiii]/\hb. Not unexpectedly, the KBSS galaxies are significantly more massive ($\sim$0.5-1.3~dex) at a given [\oiii]/\hb. Using the relations between N/S and stellar mass at low and high redshift shown in Figure~6, we can compute the expected offset in N/O at each of the [\oiii]/\hb\ levels. We find that an offset of $\sim$0.2-0.4~dex is expected based on the observed difference in galaxy masses at fixed [\oiii]/\hb, which matches the size of the O3N2 shift.

\subsection{Interpreting evolution in the N/O-$M_{*}$ relation}
Modest redshift evolution of the N/O-$M_{*}$ relation can probably best be understood as an age effect. Assuming that intermediate-mass stars are important for nitrogen production, there must be a delay before the nitrogen is released to the ISM. Galaxies of a given mass at high redshift will tend to be younger (in the sense that they are observed closer to their epoch of peak star formation) than galaxies of similar mass in the local universe \citep{Abramson16}. Therefore, they will not have reached their full N/O value yet, because latent nitrogen which will eventually be released to the ISM is still locked up in stars. Another age-related effect that can contribute to the observed evolution is the differential stellar mass loss between older and younger galaxies -- i.e., older galaxies of a given total past star formation will tend to have less remaining stellar mass than younger galaxies with the same total past star formation. Assuming the observed evolution is due to age differences in the populations, we can also explain why there is essentially no difference in the relation at high masses between the \citet{Kashino16} and local samples: at high enough mass, galaxies will tend to be mature at both $z\sim1.6$ and $z\sim0$.

To summarize, the evidence suggests that the N/O-$M_{*}$ relation is more fundamental than the N/O-O/H relation, but that it does vary slightly with redshift. However, this variation is much slower than would be expected if the N/O-O/H relationship were redshift independent. Given the likely role of inflows of pristine hydrogen in driving the evolution of the mass-metallicity relation, it would be surprising if the N/O-O/H relation held in a redshift-independent way, because O/H is affected by inflows while N/O is not.  We argue that the observed evolution in the N/O-$M_{*}$ relation can be explained as an age effect. Therefore, there is likely a still more fundamental relation between N/O, $M_{*}$, and galaxy age.

\begin{figure*}[htb]
\centering
    \includegraphics[width=.8\textwidth]{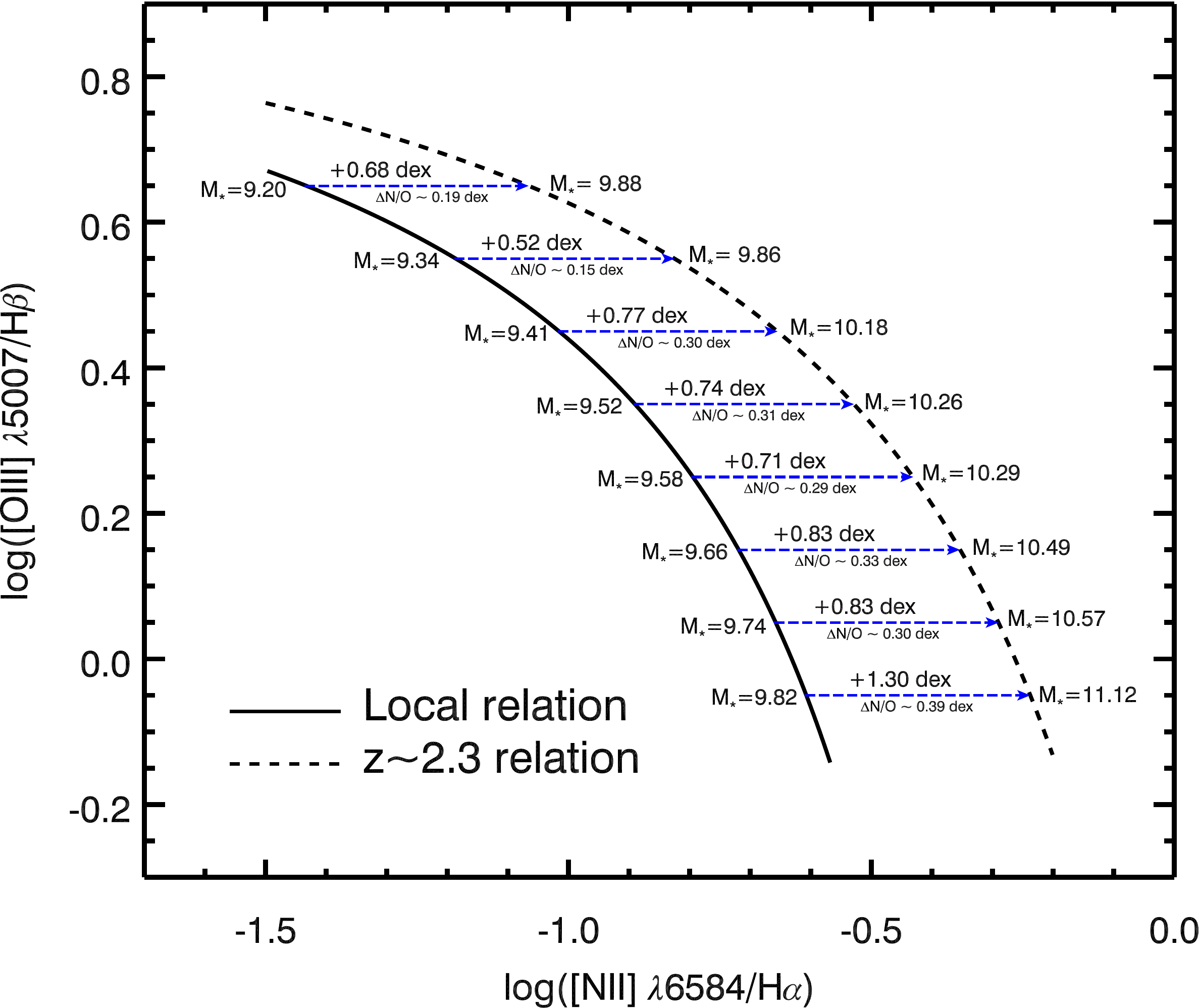} 
  \label{figure:bpt_mass}
  \caption{Using the local SDSS data and the published $z\sim2.3$ data from \citet{Steidel14}, we show the average mass offset of galaxies at different fixed [\oiii]/\hb\ values at the two redshifts. The median masses in each distribution are computed in eight bins of size 0.1 in log([\oiii]/\hb); median stellar masses at each [\oiii]/\hb\ value and redshift are shown. As expected, the $z\sim2.3$ sample is significantly more massive at fixed [\oiii]/\hb. Using the N/S versus mass relation for high redshift galaxies illustrated in Figure~6 (dotted blue line), we can predict the corresponding N/O offset at fixed [\oiii]/\hb; these predicted $\Delta$N/O values are shown, given the observed mass differential. The $z\sim2.3$ galaxies are predicted by the N/S-$M_{*}$ relations to have N/O ratios $\sim$0.2-0.4 dex higher at fixed [\oiii]/\hb, which fully accounts for the observed offset in [\nii]/\ha.}
\end{figure*}

\section{Discussion}

\subsection{Implications for metallicity measurement from strong lines at low and high redshift}
A consequence of a more fundamental N/O-$M_{*}$ relation is that metallicity diagnostics involving nitrogen will not accurately reflect a galaxy's current gas-phase metallicity, and in fact will be at least partially sensitive to its mass. Calibrations involving the N/O or N/S ratios, either implicitly (e.g., the N2  \citep{Storchi94} and O3N2 \citep{Pettini04} diagnostics) or explicitly, (e.g., \citealp{Dopita16}) should therefore be used with caution. These measures will be sensitive to galaxy mass (and age) in addition to instantaneous gas-phase metallicity.  Measurements of the MZR at high redshift (e.g., \citealp{Erb06, Mannucci09, Henry13, Steidel14, Sanders15, Faisst16b}) should be compared carefully after accounting for the possible link of nitrogen-based calibrations to stellar mass. 

Such an effect could explain why, for example, the evolution in the MZR is found to be more modest when the N2 indicator is used. \citet{Erb06} found that the $z\sim2$ MZR was $\sim$0.3~dex lower than the local relation measured by \citet{Tremonti04} if the N2 diagnostic was used on both samples, but $\sim$0.56~dex lower if the original metallicity estimates provided by Tremonti et al. were used instead. Similarly, \citet{Cullen14} derive the MZ relation at $z\sim2$ and find it offset lower than the \citet{Erb06} relation, which they attribute to the different metallicity calibrators used. 

The N/O-$M_{*}$ relation may also account for the unexpectedly tight relation between $M_{*}$ and the strong line metallicities measured in the KBSS sample, which were derived with the nitrogen-based indicators N2 and O3N2. \citet{Steidel14} note that the scatter in their derived MZ relation is actually smaller than what would be expected based on the uncertainty in the metallicity calibrations themselves, even if mass and metallicity were perfectly correlated. This observation implies a tighter relation of the observed line ratios with $M_{*}$ than with metallicity, which could be explained by a tight N/O-$M_{*}$ relation.

If, as we have argued, gas-phase metallicity is what sets the ionization parameter, then [\neiii]/[\oii], [\oiii]/[\oii], [\oiii]/\hb, or any other ratio that is primarily sensitive to ionization parameter is likely to be a good indicator of the gas-phase metallicity. This is in agreement with the conclusion reached by \citet{Jones15} based on direct metallicity estimates at $z\sim0.8$, as well as \citet{Sanders16} based on a detailed analysis of data from the MOSDEF survey. A direct metallicity calibration of the [\oiii]/[\oii] versus ([\oiii]+[\oii])/\hb\ (O32R23) diagnostic diagram, as suggested by \citet{Shapley15}, should be valid and redshift-independent, as it is effectively based on the link between ionization parameter and metallicity.

If N/O is in fact fundamentally linked to galaxy stellar mass and age, it is an interesting quantity to study at different redshifts. The differential evolution of N/O, gas-phase metallicity, and stellar mass across redshifts may provide key insights into galaxy growth and chemical enrichment over cosmic time.

\subsection{Local analogs to high redshift galaxies and the mass-metallicity-SFR relation}

We explain the offset of local high-redshift analog galaxies in O3N2 as arising from unusually high gas inflows (needed to trigger the high SFRs), which simultaneously lower their metallicity, raise their ionization parameters (and thus [\oiii]/\hb\ ratios), and leave their N/O ratios (set by stellar mass) untouched. This has the combined effect of offsetting the galaxies up and to the right in O3N2.

This scenario depends implicity on the existence of a relation between mass, metallicity, and star formation rate -- the so-called fundamental metallicity relation of galaxies (FMR, \citealp{Ellison08, Mannucci10, Lara-Lopez10,Salim14}). We make a similar argument for the offset of high redshift galaxies as for local analogs: a galaxy at $z\sim2$ has undergone a certain amount of chemical enrichment which depends principally on its stellar mass, and sets the N/O ratio, but its current (low) gas-phase metallicity is strongly affected by the inflow of pristine hydrogen gas, which also drives up its SFR. Again, we are assuming the existence of at least some mass-metallicity-SFR relation at high redshift, although making no strong claims about whether it might vary from the $z\sim0$ relation. Evidence for such a relation at high redshift does now exist \citep{Maier14,Salim15}, although it is still unclear whether it varies with redshift \citep{Sanders15}.

Three studies \citep{Wuyts14, Steidel14, Sanders15} have reported little variation of metallicity with SFR at a fixed mass in high redshift samples, which would argue against the existence of the FMR. We note, however, that all of these samples used nitrogen-based metallicity calibrations, which (owing to their closer link to mass than metallicity) will not reveal such a correlation as strongly as would a direct metallicity determination. 
 
\subsection{Interpreting the correlation between \sigsfr\ and position in O3S2}

Our primary focus up to now has been to understand the behavior of galaxies in O3N2. However, an intriguing observation arising from Figure~1 is that galaxies also segregate strongly in O3S2 space according to \sigsfr, with the most strongly star-forming galaxies falling on
the lower left side of the ionization sequence in O3S2 (lower [\sii]/\ha\ at fixed [\oiii]/\hb). This displacement strongly disfavors shocks, AGN, or significantly harder radiation sources in these galaxies, which would all be expected to cause a shift in the opposite direction in the diagram. 

Recently, \citet{Kashino16} suggested a higher ionization parameter to explain a similar shift seen in the FMOS-COSMOS data. However, the orthogonality between ionization parameter and \sigsfr\ shown in Figure~1 implies that ionization parameter is not the principle reason for the segregation according to \sigsfr\ (the same ionization parameter is found in galaxies with both high and low \sigsfr, so there must be other factors involved). It is tempting to ascribe the effect to the mass-metallicity-SFR relation, with lower [\sii]/\ha\ values reflecting a lower metallicity after an inflow of pristine gas. However, we argued previously that ionization parameter and gas-phase metallicity are closely linked; presuming this is true, it is unlikely that the segregation with \sigsfr\ arises due to gas-phase abundance variation at fixed ionization parameter. However, we do not rule it out.

Two other possibilities relate to the electron density of the gas and the size of the [\sii] emitting zone. First, we note that the electron density of the gas is weakly correlated with \sigsfr. Therefore, if [\sii] emission is significantly supressed at higher densities/pressures it might explain the correlation, but this seems unlikely given the small changes in density involved (from $\sim$30~cm$^{-3}$ on the local relation to $\sim$200~cm$^{-3}$ in the offset region with high \sigsfr). Second, we note that sulfur has a significantly lower ionization potential than nitrogen or hydrogen, meaning that the S$^{+}$ region is typically much more extended than the \hii\ region. In galaxies with very dense and widespread star formation, it is plausible that \hii\ regions merge and the S$^{+}$ zones shrink, lowering the [\sii]/\ha\ ratio. 

Currently we feel that we do not have a fully satisfying explanation for this trend, and it is worthy of further study. This offset does seem to be (perhaps marginally) apparent in high-redshift samples from MOSFIRE \citep{Shapley15} and was noted explicitly by \citet{Kashino16}. The offset in O3S2 tends to be more apparent for higher mass galaxies, which explains why it was
not seen in the \citet{Masters14} sample of low-mass galaxies from the WISP \citep{Atek10} survey. 

\section{Conclusion}

Using a large sample of star-forming galaxies from SDSS DR12, we 
explored the strong, empirical correlations between measurable
galaxy properties and position on the O3N2 and O3S2 diagrams. Our primary goal was to gain insight into the observed shift of high redshift galaxies in O3N2. We find that local galaxies falling close to the high redshift locus in O3N2 have  higher ionization parameters and \sigsfr\ values \emph{at fixed mass} in comparison with normal galaxies, which drives them up in [\oiii]/\hb, as well as higher N/O ratios \emph{at fixed ionization parameter}, which offsets them from the local sequence to the right in [\nii]/\ha. Stellar mass and N/O are observed to be distributed very similarly over the diagnostic diagrams, suggesting a close link between these quantities irrespective of the evolutionary state of the galaxy. This is in contrast to the N/O-O/H relation, which has a strong secondary dependence on SFR \citep{Perez13, Andrews13, Brown16}.

Based on these observations, we suggest that the relationship between N/O and stellar mass (N/O-$M_{*}$) is more fundamental than the relation between N/O and O/H. The stronger dependence of N/O on $M_{*}$ than O/H makes physical sense, given that the N/O ratio is the result of the integrated processing of material from prior generations of intermediate and massive stars, and is relatively independent of variations of gas-phase metallicity caused by inflows of pristine hydrogen gas or outflows of metals. Both the tightness of the N/O-$M_{*}$ relation in comparison with the correlation between N/O-O/H (e.g., \citealp{Andrews13}, Figure 14) and the absence of significant secondary dependences of N/O-$M_{*}$ on SFR (as found with the N/O-O/H relation) favor this interpretation.

Given this hypothesis, the offset in the O3N2 diagram at high redshift can be explained as follows. High redshift galaxies have much lower gas-phase metallicities and higher ionization parameters than local galaxies of the same mass (the evolution of the MZ relation), pushing them upward in [\oiii]/\hb. However, their N/O values are only modestly lower than those of local galaxies with the same mass, due to the slow evolution of the N/O-$M_{*}$ relation, and in fact are much \emph{larger} than the N/O values of local galaxies with the same [\oiii]/\hb\ ratio, which produces the observed O3N2 offset. No appeal to enhanced nitrogen enrichment of the ISM through WR-star ejecta or the effects of massive star rotation/binarity are required to explain the BPT shift under this scenario. It arises as a natural consequence of a fixed or slowly-varying relation between N/O and stellar mass in conjunction with the rapidly evolving mass-metallicity relation.

The relatively small amount of published data on the N/O ratio at high redshift suggests that N/O-$M_{*}$ does evolve modestly with redshift. We interpret this evolution as an age effect: high redshift galaxies at a fixed $M_{*}$ tend to be younger and thus have not had time to fully enrich the ISM with nitrogen from intermediate mass stars. We suggest that a more fundamental N/O-$M_{*}$-age relation is likely to exist. Further investigation of the N/O-$M_{*}$ relation and its detailed evolution with redshift would be useful both to confirm the evolution of the N/O-$M_{*}$ relation and gain insight into the causes of its variation.

\acknowledgements
D.M. acknowledges the First Carnegie Symposium in Honor of Leonard Searle for many enlightening talks and conversations. A.F. acknowledges support from the Swiss National Science Foundation. D.M. is grateful to Alaina Henry, Samir Salim, Charles Steinhardt, John Silverman, Lin Yan, Jin Koda, and Louis Abramson for helpful and enjoyable conversations. 

\bibliographystyle{apj}
\bibliography{biblio}

\end{document}